# On the influence of Langmuir wave spectra on the spectra of electromagnetic waves generated in solar plasma with double plasma frequency

Igor V. Kudryavtsev[1], T.I. Kaltman[2]

[1] *Ioffe Institute, Saint Petersburg, Russia*

[2] *Special Astrophysical Observatory of the Russian Academy of Sciences, Saint Petersburg, Russia*

igor.koudriavtsev@mail.ioffe.ru

## ABSTRACT

In this paper, we consider the spectral dependences of transverse electromagnetic waves generated in solar plasma at coalescence of Langmuir waves. It is shown that different spectra of Langmuir waves lead to characteristic types of transversal electromagnetic wave spectra, what makes it possible to diagnose the features of the spectra of Langmuir waves generated in solar plasma.

## 1 INTRODUCTION

Radio emission of solar plasma is one of the main sources of information about the physical processes acting on the Sun. This radiation can be either thermal in nature (see, for example, Zheleznyakov 1996), or non-thermal, generated by gyrosynchrotron and free-free mechanisms during the propagation of fast electrons in a magnetically active plasma (Nakariakov et al. 2018; Fleishman et al. 2018; Gary et al. 2018a,b) or generated by plasma waves (Zheleznyakov & Zaitsev 1970; Tsytovich 1977; Kontar et al. 2012). In the latter case, the generation of transverse electromagnetic waves at the electron-plasma frequency $\omega_{pe}$ can occur when Langmuir waves are scattered on plasma particles (see, for example, Tsytovich 1977). In addition, the plasma can emit radiation at frequencies of about $2\omega_{pe}$, which arises at coalescence of two Langmuir waves (Zheleznyakov & Zaitsev 1970; Tsytovich 1977). This mechanism of plasma emission is currently the accepted model for type III burst generation (Kaplan & Tsytovich 1973; Reid & Ratcliffe 2014). At the same time, Langmuir waves can be generated, for example, by beams of fast electrons accelerated during solar flares. Thus, the paper (Kudryavtsev et al., 2019) presents the results of a numerical solution of the problem of spatial propagation of fast electrons in a flare plasma, taking into account their interaction with Langmuir waves. It is shown that at the first (initial) stage of interaction, Langmuir waves are generated in a narrow range of phase velocities (or wave numbers). Then, as fast electrons propagate deeper into the plasma, a "plateau"-like distribution of the velocities of fast electrons is formed and the spectrum of excited waves expands. An asymmetric spectrum of Langmuir waves is formed (see also Kontar et al. 2012). After the formation of the "plateau"-like distribution of velocities of fast electrons, the generation of plasma waves stops until a section with a positive derivative is formed on the electron distribution function as a result of Coulomb collisions with plasma particles (see also Zheleznyakov & Zaitsev 1970), after which the Langmuir waves begin to be excited again. Thus, the dynamics of the spectrum of Langmuir oscillations will reflect the dynamics of the propagation of fast electrons that excite Langmuir turbulence. Thus, the registration of the radio emission of solar flares makes it possible to diagnose both the solar plasma itself and to determine the characteristics of high-energy electrons.

This work is devoted to the consideration of the peculiarities of the radiation of transverse electromagnetic waves at frequencies of about $2\omega_{pe}$, generated in the solar plasma via the coalescence of Langmuir waves. We will consider radio bursts with the width of the radiation spectrum $\Delta f$, which is significantly lower than the radiation frequency $f$, i.e. $\Delta f << f$.

Let us discuss in more detail the possible shape of the spectra of Langmuir waves generating type III bursts. When generating Langmuir waves as a result of the development of beam instability, some features of the spectra of these waves should be noted. These waves are generated at phase velocities lower than the speed of light *c*. At the initial stage of the development of the beam instability, an almost symmetric spectrum is formed with respect to the wave number (see, for example, Fig. 6 in Reid and Ratcliffe, 2014), which corresponds to the condition of resonance of waves and fast electrons. With the development of beam instability and further transformation of the fast electron distribution function, the Langmuir wave spectrum is formed with an extremely strong increase in the wave number and a smoother further decline, as well as with a break at the maximum of the wave number (Fig. 6 in Reid and Ratcliffe, 2014). All this makes it possible to describe the spectrum of Langmuir waves generated by fast electrons, either as a Gaussian distribution for a symmetric spectrum with respect to a certain wave number, or as an arc spectrum for an asymmetric distribution, as was done in (Willes, Robinson & Melrose 1996). Thus, in (Willes et al, 1996) the case of axial symmetry was considered, the spectrum of Langmuir waves propagating along the axis of symmetry was given as

$$W_f^l(k,\vartheta) \propto \exp\left(\frac{a(k)+b(k)\cos\vartheta}{K^2_f}\right), \quad (1)$$

where a(k) and b(k) are some functions, the choice of which makes it possible to obtain either a Gaussian distribution or an arc spectrum; the parameter $K_f$ defines the width of the wave distribution; $\vartheta$ is the angle between the wave vector **k** and the axis of symmetry. A similar expression was also used by Willes et al. (1996) for waves propagating in the opposite direction.

It should be noted here that these types of spectra (Gaussian and arc spectra) of Langmuir waves are not the only ones possible for astrophysical plasma. Thus, in (Liperovskii & Tsytovich 1970, Tsytovich 1977), the isotropic stationary spectra of Langmuir turbulence were theoretically considered and it was shown that, at a sufficiently high wave generation power, induced scattering of waves by plasma particles can transform the wave spectrum and expand it towards small wave numbers, and waves will appear with phase velocities exceeding the speed of light *c*. Otherwise, even if we take into account the absorption of waves due to collisions and the presence of fast particles in the plasma, such an expansion of the Langmuir turbulence spectrum may not occur. In the latter case, the turbulence spectrum can be formed with a smooth increase in the wavenumber to a certain characteristic value, followed by a sharper decline (see for example Fig. 4.4, curve 4 in Tsytovich 1977 and Liperovsky & Tsytovich 1970). Thus, an estimate of Langmuir spectra for individual observed solar radio bursts can also show whether this spectrum is broadened as a result of induced scattering of waves by plasma particles under these specific conditions or not.

## 2 DEPENDENCE OF THE RADIATION SPECTRUM AT THE DOUBLE PLASMA FREQUENCY ON THE LANGMUIR SPECTRUM

The expression for the spectral power of radiation of transverse electromagnetic waves formed by coalescence of Langmuir waves with the wave vectors $\mathbf{k_1}$ and $\mathbf{k_2}$, has the form (Tsytovich 1977):

$$Q_{\mathbf{k}} = \int Q_{\mathbf{k},\mathbf{k}_1,\mathbf{k}_2} W^l_{\mathbf{k}_1} W^l_{\mathbf{k}_2} d\mathbf{k}_1 \, d\mathbf{k}_2; \qquad (2)$$

$$Q_{\mathbf{k},\mathbf{k}_1,\mathbf{k}_2} = \frac{\pi\omega \left(k_1^2 - k_2^2\right)^2 [\mathbf{k}_1\mathbf{k}_2]^2}{8mnk^2 \omega_{pe} k_1^2 k_2^2} \delta(\mathbf{k} - \mathbf{k}_1 - \mathbf{k}_2) \times \\ \delta\left(\sqrt{k^2c^2 + \omega_{pe}^2} - 2\omega_{pe} - \frac{3v_{Te}^2 \left(k_1^2 + k_2^2\right)}{2\omega_{pe}}\right) \qquad (3)$$

where $\mathbf{k}$ and $\omega$ are the wave vector of the transverse electromagnetic wave and its frequency, $\omega = \sqrt{k^2c^2 + \omega_{pe}^2}$; $v_{Te} = \sqrt{k_B T_e / m_e}$, $k_B$ is the Boltzmann constant, $T_e$ is the electron temperature of the plasma; $W^l_{\mathbf{k}_1}$ is the spectral energy density of Langmuir waves.

The total power of radiation generated by merging Langmuir waves is equal to:

$$Q = \int Q_{\mathbf{k}} d\mathbf{k} = \int Q_{\mathbf{k}} k^2 dk \, d\cos\alpha \, d\phi \qquad (4)$$

where α and φ are the polar and azimuth angles of the wave vector of a transverse electromagnetic wave. In this case, the angle α is the observation angle.

The spectral power of radiation at the emission frequency $Q_\omega$ can be obtained in accordance with the expression for the total radiation power:

$$Q = \int Q_\omega(\omega, \alpha, \phi) d\omega \, d\cos\alpha \, d\phi \qquad (5)$$

Comparing expressions (4) and (5) we obtain:

$$Q_\omega(\omega, \alpha, \phi) = Q_{\mathbf{k}} k^2 \frac{dk}{d\omega} \qquad . \qquad (6)$$

In our work, we will not apply the head-on approximations, which, as noted in the work (Willes at al. 1996) can be violated at small wave numbers of Langmuir waves and their narrow spectra. We will also not use the narrow-spectrum approximation proposed in (Willes at al. 1996). The cited work also contains a scheme with the spatial orientation of the wave vectors of the merger of Langmuir waves and the generated transverse wave. In order not to limit to particular cases, we present the results of calculations according to direct expressions (2, 3), without using the simplifications usually inherent in these models, which are described in detail in (Willes at al. 1996).

Let us now consider radio emission at different spectra of Langmuir waves. Taking into account the work (Willes et al. 1996), we consider the case of axial symmetry of the radiating region and the spectrum of Langmuir waves $W^l_{\mathbf{k}_1}$ in the form:

$$W^l_{\mathbf{k}_1} = B \cdot exp\left(-\frac{(k_1 - k_0)^2}{2\sigma(k_1)^2}\right) cos^\delta(\vartheta) \text{ for } k_{min} \leq k_1 \leq k_{max} \text{ and } W^l_{\mathbf{k}_1} = 0 \text{ for others } k_1, \quad (7)$$

where B is the normalizing coefficient in accordance with the normalization condition:

$$W = \int W^l_{\mathbf{k}} d\mathbf{k} = \int_{k_{min}}^{k_{max}} \int_{-1}^{1} \int_{0}^{2\pi} W^l_{\mathbf{k}} k^2 dk \, d\cos\vartheta \, d\varphi, \qquad (8)$$

W is the total energy density of Langmuir waves; $\vartheta$ and $\varphi$ are the polar and azimuth angles of the Langmuir wave, the angle $\vartheta$ is calculated relative to the axis of symmetry of the radiating region, $\delta$ is an even positive number.

Let us first consider the case when $\sigma(k) = const$, i.e., when $\sigma$ does not depend on the Langmuir wave vector.

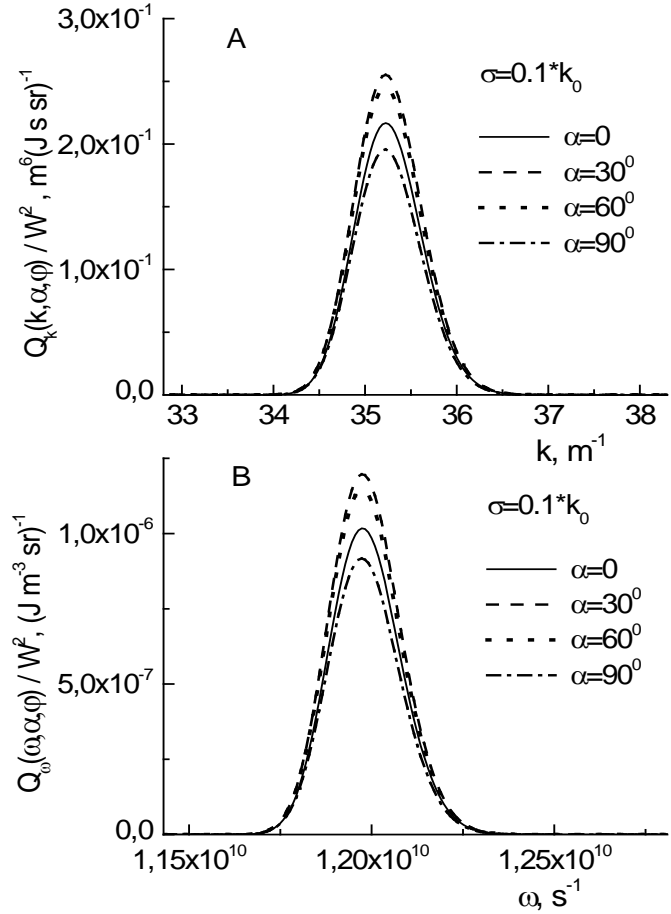

Figure 1. Spectral radiation power $Q_k$ (A) and $Q_\omega$ (B) of transverse electromagnetic waves during coalescing of Langmuir waves with spectral energy density (7) at $\sigma = 0.1 k_o$ and the plasma temperature $T_e = 3 \cdot 10^6 K$.

Fig. 1A shows the distribution of radiation ($Q_k$) over the wave number for different viewing angles $\alpha$ at $\delta=2$, $k_{max} = \omega_{pe}/(3V_{Te})$, $k_{min} = \omega_{pe}/(15V_{Te})$, and $T_e = 3 \cdot 10^6$ K, $n=10^{16}$ m$^{-3}$. These values of $k_{max}$ и $k_{min}$ are chosen for the following reasons: Langmuir waves with a value greater than $k_{max}$ are effectively absorbed by thermal electrons as a result of Landau damping, and therefore we will not consider them. Langmuir waves in plasma are usually generated by beams of superthermal electrons, so consider the region of phase velocities up to $15V_{Te}$ and assume that there are no Langmuir waves with large phase velocities (smaller wave numbers). The observation angle α is the angle between the symmetry axis of the radiating region and the vector **k**. In this case, we assume that $k_0 = (k_{min}+k_{max})/2$, and σ is less than the value of $k_0$. Fig. 1B shows the distribution of radiation ($Q_\omega$) as a function of the frequency of the transverse electromagnetic wave. The calculation results for this case are shown in Fig. 1 and 2 for different σ. At a small standard deviation (for example, σ=0.1$k_0$, Fig. 1A, B) the radiation is concentrated in a narrow frequency range. A similar type of Langmuir spectra can be generated, for example, by a fast electron beam. With increasing σ, the frequency range expands (Fig. 2A, B).

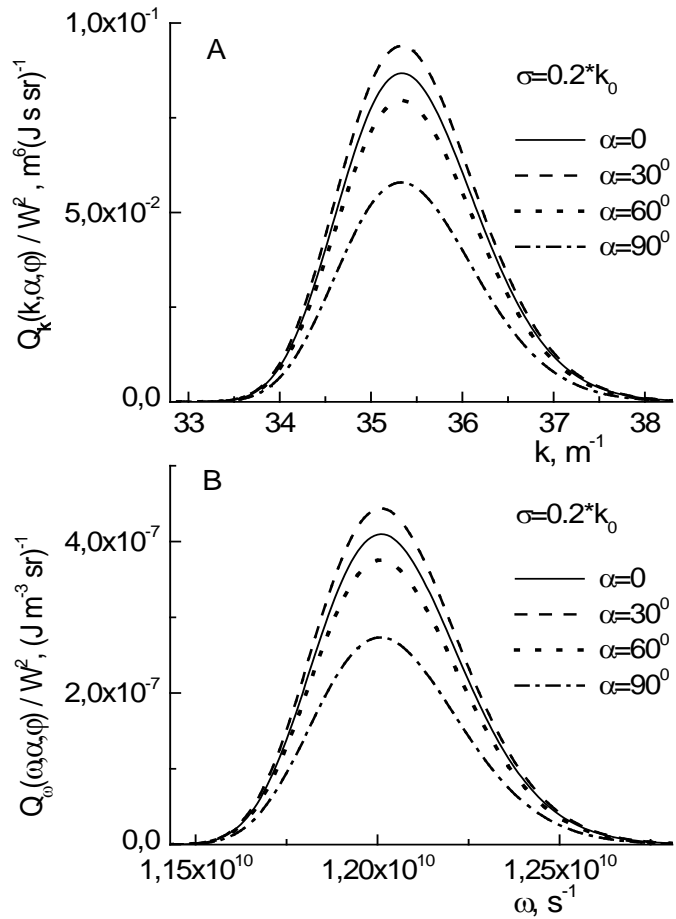

Figure 2. Spectral power of radiation $Q_k$ (A) and $Q_\omega$ (B) of transverse electromagnetic waves for coalescing of Langmuir waves with spectral energy density (7) for σ=0.2$k_o$ and the plasma temperature $T_e=3\cdot10^6$K.

Expression (3) allows to determine the range of frequencies and wavenumbers at which electromagnetic radiation is generated, i.e. for which the expression under the Delta-function vanishes. The frequency of the electromagnetic wave in the plasma is related to the wave vector by the expression $\omega = \sqrt{k^2 c^2 + \omega_{pe}^2}$, so the first term in the argument of the Delta function is the frequency of the electromagnetic wave ω. The value of this frequency ω is obtained from the condition of vanishing the argument of the Delta function in expression (3), i.e. from the law of conservation of energy: $\omega = 2\omega_{pe}(1 + 3V_{Te}^2(k_1^2 + k_2^2)/(4\omega_{pe}^2))$. Therefore, the maximum frequency of radiation $\omega_{max}$ is reached at $k_1 = k_2 = k_{max}$. For $k_{max} = \omega_{pe}/(3V_{Te})$ we obtain $\omega_{max} = 2\omega_{pe}(1 + 1/6) = 2\omega_{pe} \cdot 1.167$. The minimum frequency $\omega_{min}$ is obviously reached $\omega_{min} = 2\omega_{pe} \cdot 1.007$ at $k_1 = k_2 = k_{min}$ and at $k_{min} = \omega_{pe}/(15V_{Te})$. Thus, the frequency spectrum width of the generated radio emission $(\omega_{max} - \omega_{min})/\omega_{min}$ is about 16%. In our case, when the plasma density is $10^{16} m^{-3}$, the transverse electromagnetic waves generated in the plasma are in the frequency range from $\omega_{min} \approx 1.14 \cdot 10^{10} s^{-1}$ to $\omega_{max} \approx 1.32 \cdot 10^{10} s^{-1}$ and with the wave numbers from $\approx 32.9 \, m^{-1}$ to $\approx 39.7 \, m^{-1}$.

In the introduction, it was noted that in some cases, due to the induced scattering of Langmuir waves on plasma particles, it is possible to expand the spectrum of Langmuir waves to the region of large phase velocities exceeding the speed of light *c*, and in this case the spectrum of these waves extends from $k_{min} = 0$ to $k = k_{max}$. Then $\omega_{min} = 2\omega_{pe}$ and at $k_{max} = \omega_{pe}/(3V_{Te})$ the spectrum width of Langmuir waves $(\omega_{max} - \omega_{min})/\omega_{min}$ is about 17%. Given the above, we can conclude that if the width of the observed burst spectrum is, for example ≈10% and $k_{max} = \omega_{pe}/(3V_{Te})$, then the maximum phase velocity of Langmuir waves is ≈ $5V_{Te}$, so the expansion of the spectrum of Langmuir waves to the region of phase velocities exceeding the speed of light in solar plasma with a temperature of $10^6$-$10^7$ K does not occur.

Before considering the observational data, let us consider another type of Langmuir wave spectrum. As noted in the Introduction, the Langmuir wave spectrum can grow smoothly with an increase in the wave number to a certain value, followed by a sharp decrease. As such a type of spectra, we take the following dependence of the spectral energy density of waves on the wavenumber $k$ and on the angle $\vartheta$:

$$W^l_{\mathbf{k}_1} = A \cdot k_1^{\gamma} \cos^{\delta}(\vartheta) \quad \text{for} \quad k_{min} \leq k_1 \leq k_{max} \quad \text{and} \quad W^l_{\mathbf{k}_1} = 0 \text{ for other } k_1, \qquad (9)$$

where $\gamma$ is the spectral index, $\delta$ is an even positive number, A is a normalizing factor according to the expression (8).

So, Fig. 3A, B shows the distribution of radiation $Q_k$ and $Q_\omega$ for different viewing angles $\alpha$ at $\gamma=3$, $\delta=2$, $k_{max}=\omega_{pe}/(3V_{Te})$, $k_{min}=\omega_{pe}/(15V_{Te})$, and $T_e=3\cdot 10^6$ K, $n=10^{16}$ m$^{-3}$.

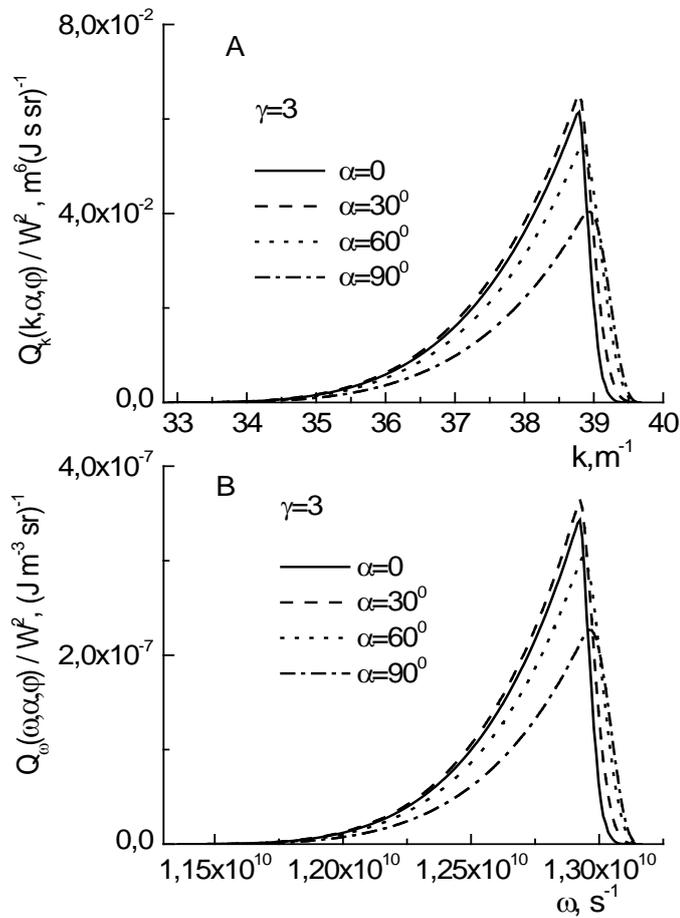

Figure 3. Spectral radiation power $Q_k$ (A) and $Q_\omega$ (B) of transverse electromagnetic waves for coalescing of Langmuir waves with spectral energy density (9) at the plasma temperature $T_e=3\cdot 10^6$ K.

As can be seen from Fig. 3A, B, the radiation spectrum in the case of Langmuir wave spectrum in the form (9) increases both with the wave number and with frequency to certain values of the wavenumber and frequency, which corresponds a discontinuity in the expression (9) used for the Langmuir spectrum.

Thus, the measurement of the frequency spectrum of radiation of electromagnetic waves generated in plasma by the merging of Langmuir waves make it possible to determine the type of Langmuir spectrum in the plasma and, therefore, provide indications for understanding the mechanism of Langmuir wave generation.

Let's compare the calculated results with the registered radio bursts. It was noted above that the width of the radio emission spectrum generated by a pairwise merger of Langmuir waves can be about 16%. Jiřička et al. (2001) give an overview of radio bursts with a fine structure in the range of 0.8-2 GHz measured in 1992-2000 with Ondrejov radio spectrograph. In particular, a review of narrowband type III bursts is given and it is noted that the typical width of these bursts is $55 - 192$ MHz, and their frequency lies in the range of $0.8 - 1.3$ GHz. Thus, the width of the spectrum of these bursts corresponds to the width of the calculated radio emission spectrum for the case of coalescence of plasma waves.

**3 RADIO BURST ON DECEMBER 1, 2004**

To make it possible to diagnose solar plasma during bursts, it is of interest to model the shape of the radio spectrum. As an example, let's take the measured instantaneous spectrum of the burst on December 1, 2004 (Huang & Tan 2012, Figure 4b – the left peak). In the cited paper, the spectra are in relative units.

Assuming that this radiation is generated by merging of Langmuir waves, let's try to determine the spectrum of Langmuir waves. Fig. 4A shows the calculated radio emission spectra consisting of the sum of the background radiation (60%) and the radiation generated by merging of Langmuir waves (40%). This and the following figures show the results of calculating for the viewing angle $0^0$. The Langmuir spectrum of the form (7) is used for calculations with plasma density and temperature of $3 \cdot 10^{15}$ m$^{-3}$ и $T_e = 6 \cdot 10^6$K, respectively, and for $k_0=0.48k_{max}+0.52k_{min}$. Curve 1 corresponds to the standard deviation $\sigma= 0.12k_0$, curve 2 corresponds to $\sigma= 0.2k_0$. As follows from this figure, the curve 1 fits the width of the measured spectrum (curve 3, data is scanned and digitized from (Huang & Tan 2012). However, the results of this simulation cannot describe the asymmetry of the measured radiation.

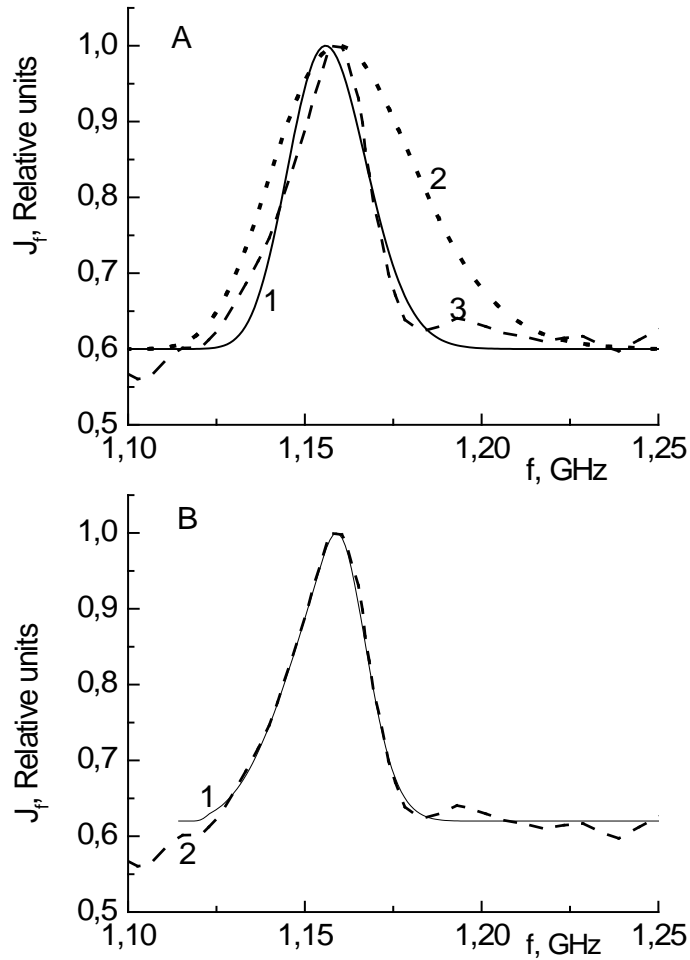

Figure 4. A − Calculated electromagnetic radiation spectra (curves 1, 2) for the Langmuir spectrum in the form (7) and the measured spectrum (curve 3) of the burst on December 1, 2004 (Huang & Tan 2012); B − Calculated electromagnetic radiation spectra (curves 1) for the asymmetric Langmuir spectrum and the measured spectrum of the burst on December 1, 2004 (Huang & Tan 2012) (curve 2). $f=\omega/(2\pi)$

The asymmetry of the measured spectrum can be the result of the asymmetry of the Langmuir spectrum with respect to $k_0$. Fig. 4B shows the calculation results for the asymmetric case (curve 1). In this case, the total radiation consists of radiation generated by the coalescence of Langmuir waves (38%) and background radiation (62%). These calculations were performed for a plasma concentration of $3.8 \cdot 10^{15}$ m$^{-3}$, a plasma temperature of $6 \cdot 10^6$ K and a Langmuir wave distribution of the form (7), but with different σ: $σ=0.3k_0$ for $k<k_0$ and $σ=0.07k_0$ for $k>k_0$, where $k_0=0.48k_{max}+0.52k_{min}$. Thus, Fig. 4B shows that the radio emission spectrum measured in (Huang & Tan 2012) (curve 2) can be described as radiation generated by the pairwise coalescing of Langmuir waves to form electromagnetic radiation at a frequency approximately twice that of the plasma frequency.

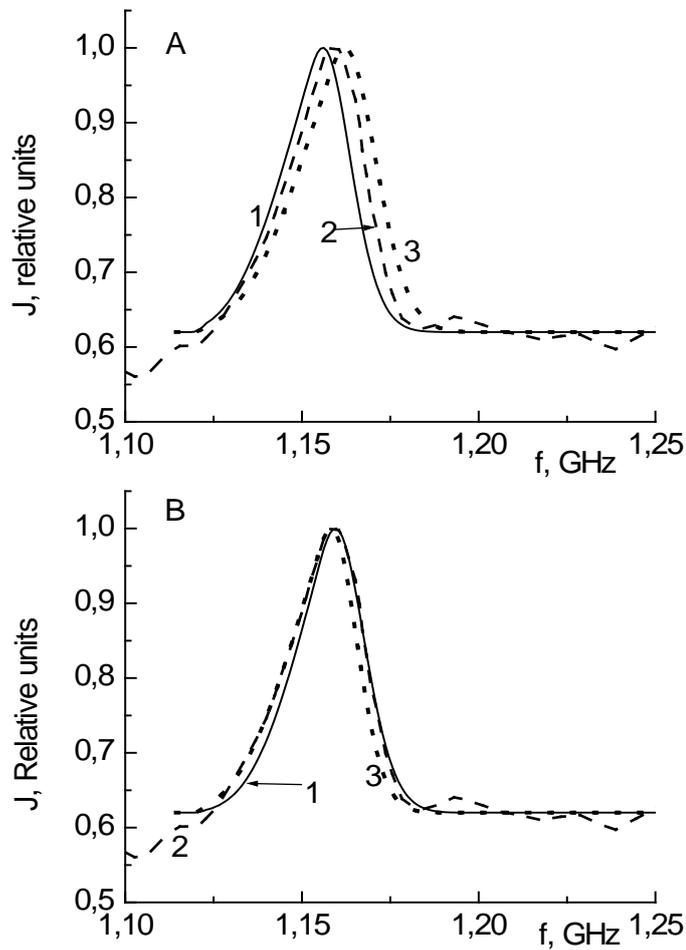

Figure 5. A- Calculated electromagnetic spectra for $k_0=0.46k_{max}+0.54k_{min}$ (curve 1), for $k_0=0.5k_{max}+0.5k_{min}$ (curve 3) and the measured spectrum of the burst (Huang & Tan 2012) on December 1, 2004 (curve 2). $f=ω/(2π)$; B- Calculated electromagnetic spectra for $σ=0.26k_0$, $k<k_0$ and for $σ=0.07k_0$, $k>k_0$, (curve 1); for $σ=0.3k_0$, $k<k_0$ and for $σ=0.06k_0$, $k>k_0$ (curve

3) and the measured spectrum of the burst (Huang & Tan 2012) on December 1, 2004 (curve 2), $k_0=0.48k_{max}+0.52k_{min}$; $n= 3.8 \cdot 10^{15} m^{-3}$. $T_e= 6 \cdot 10^6$ K.

The results of the calculations are very sensitive to changes in the parameters of the Langmuir spectrum. Thus, Fig. 5A shows the calculation results for the other two values of $k_0$ while preserving the plasma parameters and other characteristics of the Langmuir wave spectrum. Changing this parameter leads to a shift in the frequency spectrum of radiated radio waves. Changing the parameter σ also leads to a deviation of the calculated spectrum from the observed one (Fig. 5B).

In conclusion of this section, we note that the obtained spectrum of Langmuir waves (for the December 1, 2004 burst) lies in the range of phase velocities less than the speed of light *c*, which indicates that the expansion of the plasma wave spectrum to phase velocities exceeding the speed of light *c* does not occur.

## 4 CONCLUSIONS

As a result of this work, the radio emission spectra generated by coalescence of Langmuir waves were calculated in the frequency range near the double plasma frequency. It is shown that the shapes of the simulated radio emission spectra depend on the parameters of the Langmuir wave spectra. Further application of this method should make it possible to determine the nature of Langmuir waves generation from radio observations. In particular, if Langmuir waves are generated by fast electron beams at a speed of $v_0$ then these waves will occur in a narrow range of wavenumbers near $\omega_{pe}/v_0$, what will lead to a narrow radio emission spectrum. A possible Langmuir spectrum is provided, which can be used to describe the spectrum of a narrowband burst registered on December 1, 2004, published in (Huang & Tan 2012). The resulting spectrum lies in the region of phase velocities less than the speed of light, which shows that the spectrum does not expand to the region of phase velocities exceeding the speed of light *c* as a result of induced wave scattering on plasma particles. Further detailed study of the Langmuir wave spectra based on radio bursts of different widths and different radiation power should show the influence of nonlinear processes on the Langmuir turbulence spectra of different power in the solar plasma.

**ACKNOWLEDGEMENTS**

The work is supported by the RFBR grant 18-29-21016.

**DATA AVAILABILITY**

The data underlying this article will be shared on reasonable request to the corresponding author.

**REFERENCES**

Fleishman G.D., Loukitcheva M.A., Kopnina V.Yu., Nita G.M., Gary D.E., 2018, AJ, 867(1), 81

Gary D.E., Bastian T.S., Chen B., Fleishman G.D., Glesener L., 2018a, in Murphy E., ed., ASP Conf. Ser. Vol. 517, Science with a Next Generation Very Large Array, p.99

Gary D.E. et al., 2018b, AJ, 863(1), 83

Huang J., Tan B., 2012, AJ, 745, 186

Jiřička K., Karlický M., Mészárosová H., Snížek V., 2001, A&A, 375, 243

Kaplan S.A., Tsytovich V.N., 1973 in D. Ter Haar, ed., Plasma Astrophysics: International Series of Monographs in Natural Philosophy, Pergamon, p. 316

Kontar E.P., Ratcliffe H., Bian N.H., 2012, A&A, 539, 43.

Kudryavtsev I.V., Kaltman T.I., Vatagin P.V., Charikov Yu. E., 2019, Geomagnetism and Aeronomy, 59(7), 838

Liperovskii V.A., Tsytovich V.N., 1970, Soviet Physics JETP, 30(4), 682

Nakariakov V.M., Anfinogentov S., Storozhenko A.A., Kurochkin E.A., Bogod V.M., Reid H., Ratcliffe H., 2014, Res. in A&A, 14(7), 773

Sharykin I.N., Kaltman T.I., 2018, AJ, 859, 2, 154

Tsytovich V.N., 1977, Theory of Turbulent Plasma, Consultants Bureau, New York

Willes A.J., Robinson P.A, Melrose D.B., 1996, Physics of Plasmas, 3, 149.

Zheleznyakov V.V., Zaitsev V.V, 1970, Azh, 47, 60

Zhelezniakov V.V., 1996, *Radiation in Astrophysical Plasmas,* Kluwer